# Anti-positivismo, ciencias teóricas y relatividad en la Argentina de la década de 1920

# Anti-positivism, theoretical sciences and relativity in the Argentina of the 1920s

**ALEJANDRO GANGUI**

Instituto de Astronomía y Física del Espacio, Conicet y Universidad de Buenos Aires

**EDUARDO L. ORTIZ**

Imperial College, Londres

**RESUMEN**

Hacia principios de la década de 1920 es posible detectar una fractura en la percepción de la relatividad en la Argentina, caracterizada por la aparición de una serie de estudios de índole estrictamente teórica en paralelo con otra enfocada hacia la cultura general. En este trabajo tratamos de relacionar esa fractura con los avances de la crítica anti-positivista al nivel de la cultura superior en la Argentina. Esos nuevos enfoques configuraron una visión igualmente nueva de la ciencia, que cuestionaba la metodología tradicional de las ciencias experimentales y atribuía a las ciencias teóricas una posición más destacada que en el pasado. En este trabajo se hace un análisis detallado de un estudio teórico sobre la relatividad publicado en 1923 por José B. Collo y Teófilo Isnardi, dos físicos argentinos que unos años antes habían estudiado en Alemania, y que constituye una de las contribuciones emblemáticas de este nuevo enfoque.

**Palabras clave** relatividad – positivismo – Física teórica – ciencias teóricas – Einstein.

**ABSTRACT**

Observing the early years of the 1920's, it is possible to detect a fracture in the perception of relativity theory in Argentina, characterized by the publication of a series of strictly scientific studies on this theory, in parallel with presentations aimed at a general-culture level. In this work, we attempt to relate this fracture with the advances made by Anti-Positivist ideas in the higher echelons of Argentine culture. The new philosophical approach configured a new vision of science that questioned the traditional methodology of the experimental sciences and attributed to the theoretical sciences a more prominent role than they had in the past. In this work, we present a detailed account of a 1923 paper by José B. Collo and Teófilo Isnardi, two young Argentine physicists trained in Germany, which is a representative contribution to this new trend.

# INTRODUCCIÓN

Estudios sobre el espaciotiempo y la radiación comenzaron a aparecer en la Argentina hacia fines del siglo XIX, en trabajos de Valentín Balbín, Jorge Duclout y de otros autores. En la primera década del siglo XX se publicaron también notas sobre la materia y la radiación y sobre la nueva teoría de la relatividad. Esta última parecía ofrecer una visión unificada de la materia y la energía e introducía concepciones nuevas sobre la estructura del espacio físico. Esos estudios tenían un carácter informativo serio y se difundieron en medios de prestigio científico local: revistas universitarias, revistas de ingeniería, y en los *Anales* de la Sociedad Científica Argentina. Sus autores eran, principalmente, miembros acreditados del sector académico; argentinos unos, extranjeros otros.

A partir de las observaciones astronómicas de Arthur S. Eddington, en 1919, que parecían comprobar la veracidad de la teoría de Einstein, los trabajos relacionados con esa teoría se multiplicaron. En este punto del proceso de introducción de la teoría de la relatividad en la Argentina se produjo una bifurcación de intereses, que no es tampoco ajena al proceso de introducción de la relatividad en otros países de la América Latina.

Por una parte, y como había ocurrido con otras grandes teorías en el pasado- el transformismo darwinista, por ejemplo- se pensaba en esos años que la teoría de la relatividad podría abrir panoramas que iban más allá de la estricta innovación científica, y que podrían afectar nuestra concepción de la naturaleza. A esta vertiente pertenece un grupo de trabajos, notas y conferencias, entre las que se destacan las del erudito poeta Leopoldo Lugones[i]. En ellas se tocaron problemas con un contenido cultural y a veces filosófico, en las que el nivel científico fue muy variable. De estos trabajos se hicieron eco publicaciones periódicas de un carácter más amplio que las estrictamente científicas. Por ejemplo, prestigiosas revistas culturales, como la *Revista de Filosofía*, lo mismo que las ediciones dominicales de los principales periódicos, particularmente *La Nación* y *La Prensa* de Buenos Aires, entonces decanos de la prensa argentina.

En paralelo con ese creciente aporte de explicaciones de la teoría a un nivel popular culto, hacia principios de la década de 1920 se robusteció una vertiente *científica teórica* que, con mayor o menor suerte, pretendió explicar las teorías de Einstein con un mínimo de concesiones a las dificultades que presentaba su tecnicismo original. Esos trabajos, que pertenecen a un interesante intento de aproximación a las ciencias teóricas, pretendían ser claros, pero no eran de lectura fácil. Debido a su naturaleza, y al público al que estaban enfocados, esos estudios buscaron nuevamente el amparo de revistas científicas, o el de aquellas que aspiraban a cubrir parcelas muy cercanas a la ciencia como, por ejemplo, las de ingeniería.

En una serie de trabajos muy recientes[ii] se ha intentado relacionar el interés de esos años por la teoría de la relatividad con un interesante giro hacia las ciencias teóricas que tuvo lugar dentro de las capas más aquilatadas de la cultura argentina y que se relaciona con la llamada polémica anti-positivista.

Entre las corrientes más significativas de esa polémica cabe citar la postura *epistemológica* que encabezaban los filósofos Coriolano Alberini y Alejandro Korn en la Facultad de Filosofía de la Universidad de Buenos Aires, donde iniciaron una dura y sistemática campaña contra el llamado positivismo argentino, que había dominado en los círculos intelectuales locales desde aproximadamente la última década del siglo diecinueve. Hemos usado *positivismo* como una etiqueta conveniente para indicar una forma local del positivismo francés a la que se incorporaron elementos provenientes de doctrinas asimiladas anteriormente en la Argentina, del cientificismo y del pragmatismo.

Las nuevas tendencias filosóficas de corte anti-positivista comenzaron a desarrollarse en la Argentina hacia mediados de la década de 1910 y culminaron a principios de la de 1920, cuando virtualmente habían adquirido ya control de la escena cultural local y de la enseñanza universitaria de la filosofía. El impacto de esta polémica sobre el concepto contemporáneo de ciencia, aunque poco estudiado, fue considerable y se hizo visible a través de un marcado giro del interés hacia el estudio de las ciencias teóricas. Los problemas contemporáneos de la matemática pura[iii] y la teoría de la relatividad fueron dos áreas de particular impacto.

La nueva percepción de la ciencia que apoyaban las corrientes anti-positivistas permeó hacia las instituciones nacionales encargadas de dictar las políticas científicas y, hacia principios de la década de 1920, precipitó un cambio importante que se caracteriza por una desaceleración del antiguo soporte dado por el estado a las ciencias experimentales y un nuevo interés por las ciencias teóricas[iv]. Efectivamente, hacia 1920 comenzó a debilitarse el apoyo a los grandes proyectos nacionales de ciencia experimental que habían sido iniciados a principios de siglo en Buenos Aires, Córdoba y La Plata[v].

Dentro de las diversas manifestaciones que adquirió ese interés nuevo por las ciencias teóricas podría mencionarse la decisión de invitar al joven y brillante matemático puro español Julio Rey Pastor, graduado en Alemania, y luego al físico teórico Albert Einstein, a visitar la Argentina y disertar sobre temas avanzados de las ciencias teóricas[vi]. Lo mismo puede decirse del pedido de autorización que la *Revista del Centro de Estudiantes de Ingeniería* hizo a Einstein para publicar en sus páginas una traducción de su memoria fundamental sobre la teoría de la relatividad general, originalmente publicada en *Annalen der Physik* en 1916.

Dentro de esa nueva atmósfera cultural se sitúa un interesante estudio teórico sobre la teoría de la relatividad escrito por dos jóvenes físicos argentinos, José B. Collo y Teófilo Isnardi que, con anterioridad, habían hecho estudios avanzados de física en Alemania. Debido a la calidad de su factura este trabajo es paradigmático, dentro de las ciencias físicas, de un nuevo e importante giro intelectual hacia las ciencias teóricas al que acabamos de aludir y que se operaba en la Argentina de esos años. El trabajo de Collo e Isnardi incluía una tercera parte, escrita por el astrónomo argentino Félix Aguilar[vii], entrenado también en Europa, donde discutió las implicaciones astronómicas de la relatividad. De este último estudio, que hemos analizado separadamente[viii], no nos ocuparemos en este trabajo.

Esta nota está dedicada, precisamente, a hacer un análisis puntual de la memoria de Collo e Isnardi, intentando situarla más precisamente en el contexto intelectual de su época. Si bien ese trabajo ha sido citado con frecuencia en la literatura sobre la relatividad en la Argentina, no lo ha sido siempre con un conocimiento claro de su contenido. No faltan en esa literatura referencias en las que se critica a aquella memoria, por ejemplo, por errores y ligerezas en su manejo del cálculo tensorial, cuando los autores indican explícitamente que no van a abordar el cálculo tensorial en su estudio y, efectivamente, ese cálculo no aparece jamás en ella. Lamentablemente éste y otros puntos de similar carácter sobre el trabajo de Collo e Isnardi han permeado a la literatura secundaria sobre la relatividad en la Argentina, pasando de un autor a otro por un período que se extiende hasta el presente, justificando un análisis más preciso de esa interesante memoria.

**Una breve noticia acerca de las publicaciones sobre la teoría de la relatividad anteriores a 1922**

Entre la multitud de trabajos publicados en la Argentina sobre temas relacionados con la relatividad entre 1909 y 1922 se destacan el del matemático italiano Ugo Broggi (1880-1965), aparecido en 1909, que es un estudio de la materia, la radiación y el tiempo con un contenido científico preciso[ix]. Su autor se había doctorado con una tesis sobre temas de probabilidad en Alemania bajo la supervisión de David Hilbert. Cuando apareció aquel trabajo Broggi era profesor contratado en la Universidad de La Plata, universidad que había sido creada en 1905 en un período de fuerte influencia positivista, y que se caracterizó por su consistente estímulo al desarrollo de las ciencias experimentales[x].

Un año más tarde el conocido físico-matemático y senador italiano Vito Volterra visitó Argentina como delegado oficial de Italia a las celebraciones del primer centenario de la independencia argentina. Una vez en Buenos Aires fue invitado a pronunciar una conferencia ante la Sociedad Científica Argentina, una institución fundada en 1872 en Buenos Aires que, por un período considerable de tiempo, fue la caja de resonancia de las novedades e intereses científicos de esa ciudad. En su conferencia Volterra se ocupó del espaciotiempo y la materia e hizo referencia explícita al trabajo de Einstein[xi].

Entre 1910 y 1915 el físico-matemático francés Camilo Meyer (1854-1918), condiscípulo y amigo de Henri Poincaré, dictó una serie de cursos libres sobre física-matemática en la Universidad de Buenos Aires y en el último de ellos, que ha sido publicado, introdujo por primera vez en América Latina una exposición detallada de la teoría cuántica[xii]. En ese curso hizo referencias explícitas a trabajos de Einstein, pero no a aquellos que se relacionan directamente con la teoría de la relatividad.

A fines de la década de 1900 el prestigioso astrónomo estadounidense Charles Dillon Perrine (1867-1951), del Observatorio Lick de California, fue contratado como director del Observatorio de Córdoba, el observatorio nacional de la Argentina, creado a comienzos de la década de 1870 y re-activado hacia 1910 dentro del nuevo esquema de orientación positivista que propiciaba el desarrollo de las ciencias experimentales. Perrine renovó los talleres de óptica y las facilidades de observación de esa institución y en 1912, respondiendo a un pedido de colegas suyos en Alemania asociados con Einstein, inició una serie de observaciones durante eclipses de Sol tendientes a detectar una posible deflexión de la luz. Esas observaciones comenzaron en Brasil en 1912 y continuaron luego en Rusia y Venezuela pero, debido a condiciones meteorológicas desfavorables, Perrine y sus colaboradores no pudieron ofrecer una respuesta concluyente[xiii].

A principios de la década de 1910 la Universidad de La Plata contrató al físico Jacobo Juan Laub[xiv] (1884-1962) como profesor de física. A partir de 1912 Laub comenzó a publicar en la Argentina una serie de trabajos sobre la teoría de la relatividad en *Anales* de la Sociedad Científica. Laub había tenido un contacto personal con Einstein[xv] y había colaborado con él en trabajos de investigación sobre la teoría de la relatividad[xvi], sobre los que volveremos más adelante. Las suyas fueron las contribuciones de un científico que conocía las ideas de Einstein y su teoría en detalle. Sus notas sobre relatividad comenzaron en revistas científicas de prestigio pero, a medida que la década avanzaba y el interés público por la relatividad crecía, se movieron hacia calificados medios de difusión general, por ejemplo la *Revista de Filosofía* que entonces dirigía el prestigioso médico y filósofo positivista José Ingenieros, pero que estaba orientada hacia un público con un interés cultural mucho más amplio que el de la física teórica.

Las conferencias universitarias sobre relatividad del físico español Blas Cabrera durante su visita a la Argentina en 1920[xvii] y el trabajo del astrónomo Jesuita José Ubach (también español, profesor en el Colegio del Salvador de Buenos Aires) sobre los resultados de la expedición astronómica de 1919[xviii] fueron contribuciones detalladas, orientadas a los intereses de un público culto, que apuntan ya hacia la vertiente técnica del trabajo de Collo, Isnardi y Aguilar. Otros físicos, astrónomos o ingenieros contribuyeron también con trabajos que sugieren familiaridad con la obra de divulgación de Einstein[xix], traducida al castellano en 1921[xx].

**Los físicos José B. Collo y Teófilo Isnardi**

José B. Collo, cuyo padre era diseñador de jardines, nació en Córdoba en 1887 y más tarde su familia se mudó a la provincia de Santiago del Estero. Allí una beca privada le permitió trasladarse a la ciudad de La Plata para continuar allí sus estudios. Teófilo Isnardi nació en Buenos Aires en 1890 pero hizo sus estudios secundarios en La Plata, donde su padre dirigía obras de ingeniería. Collo e Isnardi inicialmente ingresaron a la Universidad de La Plata como estudiantes de ingeniería, pero al llegar al tercer año, poco después de que el físico alemán Emil Bose fuera invitado a organizar un Instituto de Física[xxi] en esa universidad pasaron a la nueva carrera de Física, egresando en 1911. En 1912 obtuvieron su doctorado en física trabajando en temas experimentales sugeridos por Bose. En paralelo con sus estudios de Física Collo e Isnardi hicieron estudios de pedagogía, que entonces era una de las áreas de mayor modernidad en la Universidad de La Plata.

Luego de una visita que Walter Nernst hizo a la Argentina en 1912 ambos graduados recibieron una beca externa que les permitió trasladarse a Berlín; allí estudiaron bajo la supervisión de Nernst y de Max Planck. Más tarde Isnardi trabó relación con Einstein en París.

A su regreso a la Argentina ambos físicos se reincorporaron a la Universidad de La Plata y a la Escuela Naval, donde enseñaban ya desde antes de su partida al exterior. Una vez terminada la guerra, en 1920, trataron de obtener apoyo oficial para continuar perfeccionándose en física en Alemania, pero no tuvieron éxito. Más tarde ambos pasaron a la Facultad de Ciencias de la Universidad de Buenos Aires donde tuvieron a su cargo el dictado de los cursos de física general. Isnardi enseñó también físico-química y, luego de crearse la Licenciatura en Física, en 1926, fue encargado de dictar los cursos de física-matemática; fue también designado director del nuevo Departamento de Física de esa facultad. Collo, por su parte, se orientó hacia la mecánica teórica, especializándose en problemas avanzados de balística exterior, mecánica de fluidos y estabilidad mecánica.

Además de la Escuela Naval y las universidades del área del Río de la Plata, estos dos excelentes pedagogos tuvieron una extensa actuación como profesores de física en la Escuela Militar y, en la década de 1930, también en la recientemente creada Escuela Superior Técnica del Ejercito[xxii] lo que les abrió sólidos canales de comunicación con las fuerzas armadas, las que tuvieron una posición hegemónica en la vida política de la Argentina durante el medio siglo que va de 1930 a 1983. Collo e Isnardi escribieron un tratado de física general para los estudiantes de la Escuela Naval que se utilizó también en las universidades. En la segunda mitad de la década de 1950, luego de la caída del gobierno del general Juan D. Perón, ambos físicos participaron en la reorganización de la Comisión Nacional de Energía Atómica, de la que Isnardi fue designado director. Isnardi falleció en Buenos Aires en 1966 y Collo dos años más tarde.

**La vertiente teórica en la física**

En 1922-23 la Sociedad Científica Argentina, que con anterioridad había publicado en las páginas de sus *Anales* notas que indistintamente cuestionaban o favorecían la teoría de la relatividad, organizó un ciclo de conferencias sobre esa teoría[xxiii]. Hemos señalado más atrás que el comienzo de la década de 1920 coincide con una crisis severa del positivismo que, en una buena parte de los dos últimos decenios, había dinamizado considerablemente las ciencias experimentales, favoreciendo la creación de laboratorios experimentales, primero en la Universidad de Buenos Aires[xxiv], luego en la de La Plata, y finalmente en el Observatorio de Córdoba. La corriente de pensamiento anti-positivista no desestimaba en principio la importancia del conocimiento experimental, pero favorecía decididamente el avance de las ciencias teóricas. Fue precisamente en este período, la primera mitad de la década de 1920, cuando la matemática, y más tarde la física teórica, comenzaron a desarrollarse con cierto vigor en la Argentina[xxv].

El propósito de las conferencias organizadas por la Sociedad Científica era contribuir a retornar el debate sobre la relatividad al ámbito estrictamente científico y enfocarlo desde el punto de vista de la teoría. Ese debate, como hemos indicado más atrás, tenía ya una vertiente de interés cultural general que corría por un cauce paralelo.

Para el dictado de esas conferencias aquella institución invitó a los principales físicos y matemáticos de que entonces disponía la Argentina: los matemáticos Broggi y Rey Pastor, el físico-químico Horacio Damianovich (un discípulo de Meyer) y el físico alemán Ricardo Gans (sucesor de Bose en la dirección del Instituto de Física de la Universidad de La Plata hasta 1925, cuando retornó a Alemania). También se invitó a participar a los físicos Collo e Isnardi y al astrónomo Aguilar, tres de los investigadores argentinos jóvenes más destacados en ese momento en el campo de las ciencias exactas. Los dos primeros contribuyeron con un análisis de la teoría de la relatividad restringida y de la general, respectivamente, mientras que Aguilar se ocupó de la reciente verificación astronómica.

En paralelo con el debate anti-positivista, la Argentina vivía en esos años un proceso de modernización cultural y política que había alcanzado también a las universidades a través de la Reforma Universitaria de 1918. Con ella los estudios científicos superiores recibieron un estimulo considerable. Se crearon cátedras y carreras en disciplinas nuevas a la vez que se actualizaron los planes de estudios de otras. Para implementar esos cambios se hizo necesario incorporar nuevamente a profesores extranjeros de gran prestigio. Entre ellos se destacan el fisiólogo y pacifista alemán Jorge Federico Nicolai y el economista Alfonso Goldschmidt. También se promovió la creación de centros de excelencia, como el de fisiología liderado por el argentino Bernardo A. Houssay, que más tarde alcanzó reconocimiento internacional. Dentro de ese mismo proceso, en 1922 se incorporó a Rey Pastor definitivamente a la Universidad de Buenos Aires, lo que hizo posible el dictado de cursos a un nivel considerablemente más avanzado que en el pasado inmediato. En esos cursos se discutieron también temas de interés para la comprensión de la matemática de la teoría de la relatividad generalizada y, más generalmente, se crearon condiciones favorables para el surgimiento de la física matemática y, más tarde, de la física teórica.

Aparentemente, debido a que la mayoría de los conferenciantes había publicado sus ideas sobre la relatividad con anterioridad en otras revistas, *Anales* no publicó los textos de las conferencias dictadas en la Sociedad Científica. Aguilar, Collo e Isnardi, que no habían publicado su trabajo, recurrieron a una revista profesional: el *Boletín del Centro Naval*, editado por la institución social de los oficiales de la marina de guerra.

En esos años ese *Boletín* había dado cabida a notas sobre temas de electricidad, telecomunicaciones, aviación, química, meteorología y astronomía, pero en general esos artículos tenían una referencia directa a cuestiones navales. El artículo de Aguilar, Collo e Isnardi sobre la teoría de la relatividad sin duda ensanchaba el abanico de intereses del *Boletín* como ninguno antes que él (en la introducción a su trabajo los autores hicieron referencia a este hecho) pero, como hemos señalado más atrás, el profesorado de Collo e Isnardi en la Escuela Naval es posiblemente una razón de mayor peso para esa elección. Para ellos la Escuela Naval era una fuente estable e importante de recursos económicos.

Justamente en esos años Collo e Isnardi habían comenzado a desplazar el centro de gravedad de sus actividades docentes fuera de la Universidad de La Plata, buscando perspectivas nuevas en una combinación de tareas docentes compartidas entre aquella, la Escuela Naval y la Universidad de Buenos Aires. También habían comenzado a mover sus intereses científicos desde la física experimental hacia aspectos teóricos de esa ciencia.

Los tres autores declararon que su artículo había sido redactado con la expresa idea de que no fuese ni excesivamente técnico, ni que tomase la forma de los típicos "resúmenes de divulgación". En la Introducción[xxvi] los autores precisaron que sus contribuciones eran exposiciones sintéticas "para quienes sin ser especialistas poseen conocimientos suficientes para poder interesarse por algunos detalles del desarrollo de la teoría". Efectivamente, éste no es un trabajo científico original. En cambio, es un artículo serio, bien armado, aunque con concesiones al complejo aparato matemático que requería la teoría de la relatividad, sobre todo en la formulación generalizada. En el lugar adecuado Isnardi hizo explícita esa limitación en su trabajo y remitió al lector a fuentes bibliográficas precisas. Aunque consciente de que era un camino más largo el autor[xxvii] prefirió seguir la ruta indicada por Einstein en la obra conjunta de Lorentz, Einstein y Minkowski[xxviii].

Sin embargo, a pesar de sus esfuerzos el trabajo no es de fácil lectura, difícilmente podría cumplir el cometido de ser de alguna utilidad para "nuestros marinos", como los autores habían sugerido en la introducción. En cambio, podía dar a aquellos una idea clara de que los dos profesores de física contratados por la Escuela Naval eran personas excepcionalmente idóneas y que seguían su materia con asiduidad.

Como hemos señalado en otros trabajos[xxix], esos estudios contribuyeron a crear un clima favorable para la recepción de la teoría de la relatividad en la Argentina a un nivel científico más satisfactorio que el que ofrecían las obras generales de divulgación que proliferaban en diferentes idiomas, incluso en castellano.

Muy poco después, en 1926 y dentro de la misma línea de intereses teóricos, apareció la primera nota original sobre un tema de relatividad, que fue escrita por un joven graduado de la Plata, Enrique Loedel Palumbo, alumno de Gans. Loedel Palumbo había nacido en Uruguay y se educó en la Universidad de La Plata, radicándose luego en la Argentina. En una discusión organizada por la Academia de Ciencias en Buenos Aires en 1925[xxx] tuvo oportunidad de comunicar personalmente su trabajo a Einstein y luego lo publicó en Alemania[xxxi] y en la Argentina.

En las secciones que siguen nos ocuparemos de estudiar, con un cierto detalle, el contenido del trabajo conjunto de Collo e Isnardi sobre la teoría de la relatividad, su relevancia, su originalidad y las fuentes bibliográficas que les sirvieron de apoyo. El tercer artículo de la serie publicada en el *Boletín del Centro Naval*, escrito por Aguilar, que versa sobre la verificación astronómica de la relatividad general como una nueva teoría de la gravitación, ha sido considerado en otro trabajo de estos mismos autores[xxxii].

**La relatividad restringida**

Los trabajos de Collo, Isnardi y Aguilar aparecieron en tres números consecutivos del volumen 41, a partir del número 442, de septiembre-octubre de 1923. En el primero de ellos se publicó la nota de Collo que cubre veintiún páginas[xxxiii]. Collo se ocupó de explicar los *preliminares* de la teoría especial (o restringida) de la relatividad y comenzó su trabajo analizando el proceso de abstracción que condujo a Galileo a formular las leyes de la mecánica, completadas luego por Newton y, finalmente, revisitadas por Einstein con el aporte de nuevas ideas acerca del tiempo y la simultaneidad, que le sirvieron para formular los postulados de la relatividad especial. Esta sección concluye con un estudio de las transformaciones introducidas por Hendrik Antoon Lorentz y la representación geométrica del espaciotiempo debida a Hermann Minkowski[xxxiv], temas que Isnardi luego retoma y desarrolla en relación con la teoría generalizada.

Collo agudamente señala que la evolución de la física muestra diferentes escalas de dimensión que van de lo terrestre a lo planetario y finalmente a lo cósmico, y que las primeras transiciones son visibles en la problemática considerada por Galileo y Newton. A continuación define la mecánica, el estudio del movimiento de los cuerpos, y subraya las observaciones sobre las que se edificó esa ciencia, insistiendo en que sus postulados y principios básicos se dedujeron a partir de abstracciones, de cuya formulación matemática se deducen consecuencias que deben ser sometidas al veredicto de la experiencia. Es decir, una construcción abstracta que debe ser finalmente validada por la experiencia.

Collo enuncia la ley de inercia de los cuerpos (la primera ley de Newton de la mecánica clásica), también conocida como la ley de inercia de Galileo, como uno de esos principios fundacionales y luego se ocupa de la segunda ley de Newton, otro principio básico en el que se trata del efecto de una fuerza externa aplicada sobre un cuerpo y capaz de modificar, con el transcurso del tiempo, su cantidad de movimiento. Sus citas son el reciente e importante libro histórico-crítico de la evolución de la mecánica de Mach[xxxv], que aparece citado ya en la primera página de su trabajo y que continúa orientando la exposición, y el libro de Max von Laue[xxxvi].

Collo se refiere inmediatamente a las críticas que se han encontrado a la obra de Newton, tanto en lo que concierne a sus axiomas principales como a sus conceptos sobre el espacio y el tiempo. Una de estas objeciones tiene que ver con la necesidad newtoniana del espacio absoluto. Recordemos que, por una parte, la expresión matemática de las leyes de Newton requiere definir un sistema de coordenadas y que la mecánica pretende, a la vez, formular sus leyes en la forma más simple posible. En el caso de las fuerzas que actúan sobre cuerpos ubicados en un sistema rotante, como por ejemplo la Tierra, se hizo necesario introducir en el esquema de Newton ciertas fuerzas no inerciales, como la fuerza centrífuga, cuya descripción requiere interpretar el movimiento con respecto a un espacio absoluto.

Collo menciona luego la conocida experiencia del péndulo de Foucault como clave, según la teoría de Newton, para constatar el movimiento de rotación de la Tierra en el espacio absoluto de las estrellas lejanas. Luego vuelve a considerar las dos primeras leyes de Newton, escribiendo la segunda explícitamente para un sistema de coordenadas inercial (donde vale la ley de inercia de Galileo) y se pregunta si esa expresión será solo posible en un único sistema inercial o si, por el contrario, habrá más de un referencial en donde tenga validez. Demuestra entonces que en todo sistema de coordenadas, dotado de un movimiento rectilíneo y uniforme (es decir, sin aceleraciones) con respecto al primero, valdrá la misma expresión para la segunda ley. Su principal referencia es el libro de mecánica de Max Planck[xxxvii] autor por el que Collo, que había sido su alumno, conservó especial respeto a lo largo de su carrera.

En otra palabras, Collo muestra que "existen infinitos sistemas inerciales (animados de traslaciones uniformes uno respecto de los otros) y entre ellos no existe ninguno privilegiado para el estudio de los fenómenos mecánicos". Este enunciado, conocido como el *principio de relatividad para la mecánica clásica* nos dice que un observador confinado a un sistema que se mueve uniformemente (por ejemplo, un vagón de ferrocarril sin ventanas que se mueve con movimiento rectilíneo y uniforme) no podrá jamás decidir, por medio de experimentos mecánicos realizados en su sistema, cual es su propio estado de movimiento. Aun si existiese un espacio absoluto como el propuesto en la teoría newtoniana, nunca se podría afirmar que un punto del universo se halla en reposo absoluto.

En este punto de su exposición Collo señala que hasta ahora no ha introducido conceptos nuevos, y pasa a considerar la posición de los fenómenos electromagnéticos en esta teoría. Si bien hasta ahora no ha sido posible descubrir el reposo absoluto mediante experimentos mecánicos, se pregunta si "¿no será posible constatar el reposo absoluto por experiencias electromagnéticas (ópticas)?", a lo que responde, adelantando resultados que ha de desarrollar en las páginas siguientes, que "a raíz del fracaso de las tentativas hechas con este fin ha nacido la teoría de la relatividad".

Pasando revista a las diferentes concepciones del éter, el supuesto medio por el cual se propagaría la luz, Collo señala que esa hipótesis ha dado lugar a opiniones contradictorias. Hacia 1818 Augustin-Jean Fresnel, y más tarde, en 1851, Armand Fizeau, imaginaban que el éter es parcialmente arrastrado por los cuerpos mientras que Heinrich Hertz, hacia 1890, adaptando un modelo de George Stockes de 1845, suponía que el éter era efectivamente arrastrado por los cuerpos en movimiento. Al giro del siglo, Lorentz postuló que el éter es completamente inmóvil, es decir que no es arrastrado por los cuerpos y, aunque lo supone rígido, lo cree incapaz de presentar resistencia alguna al movimiento. La referencia principal de Collo en esta discusión es el libro del último autor mencionado[xxxviii].

Con palabras claras y esquemas detallados Collo describe las delicadas experiencias ópticas realizadas por Fresnel y Fizeau para estudiar el comportamiento del éter; su referencia aquí es el libro de Werner Bloch[xxxix]. Finalmente, Collo describe otro grupo importante de experiencias, basadas nuevamente en los fenómenos de interferencia de la luz y llevadas a cabo por una cadena de investigadores: Michelson en 1881, Michelson y Morley en 1887, y Morley y Miller en 1904. Mientras que Fizeau estudiaba el posible efecto de arrastre de la velocidad de la luz utilizando una corriente de agua a una velocidad relativamente baja, los nuevos experimentadores estudiaron el efecto de arrastre, o viento de éter, que podría inducir la velocidad de traslación de la Tierra alrededor del Sol, que es mucho más significativa que las velocidades de comparación utilizadas hasta entonces. Sin embargo, esas experiencias, en las que se introdujo un elemento de naturaleza astronómica, no acusaron "nunca el efecto esperado".

Luego de comentar brevemente algunas interpretaciones alternativas de los resultados de experiencias interferométricas de los autores recién citados, entre ellas la que propuso Walther Ritz, y la introducción de la llamada contracción de Lorentz-FitzGerald, Collo señala que si bien explican algunos resultados, a la vez introducen elementos que conducen a contradicciones.

En este punto comienza a presentar las novedades propuestas por Einstein en 1905, señalando que en su enfoque del problema se hace un análisis más profundo y detenido tanto de los conceptos fundamentales de la mecánica como de los de espacio y tiempo. Esos estudios llevaron a Einstein a desechar la idea de un espacio absoluto y a postular que el principio de relatividad, aceptado ya en la mecánica, vale también para los fenómenos electromagnéticos y, en particular, para los de la óptica; nos dice Collo que "en todos los sistemas de coordenadas respecto de los cuales valen las ecuaciones de la mecánica valen las mismas leyes electromagnéticas y ópticas"[xl].

Llegado a este punto Collo sigue de cerca el texto de Einstein de 1917[xli], y de él se vale para repasar las nociones de reposo y movimiento relativo, de simultaneidad y el concepto, que ahora no es absoluto, de tiempo. Toma las palabras de Einstein para convencer al lector de que el concepto de simultaneidad, por ejemplo entre dos relojes, es algo sutil, pues "nada puede asegurarnos que un reloj siga teniendo la misma marcha cuando se lo lleve a otro punto (los relojes de péndulo, por ejemplo, andan más rápidamente cerca del Polo que en el Ecuador)"[xlii]. Plantea luego los dos axiomas de la relatividad restringida: el principio de relatividad, ya mencionado anteriormente, y la constancia de la velocidad de la luz en el vacío. Este último afirma que "todo rayo luminoso se propaga respecto de cualquier sistema inercial de coordenadas con la misma velocidad, independientemente de si el cuerpo que lo emite está en movimiento o en reposo".

Luego se ocupa de formular una definición axiomática de la teoría de la relatividad y de demostrar que esos axiomas no contradicen los resultados de la experiencia y ofrece ejemplos que lo llevan a mostrar la "relatividad del tiempo y la longitud". Para ello se vale del típico ejemplo de los haces de luz emitidos en el interior de un vagón de ferrocarril en movimiento relativo a la estación. Al cabo de unos pocos pasos deduce que acontecimientos simultáneos para un sistema de coordenadas no lo serán para otro sistema, y por lo tanto enuncia que "podremos definir un tiempo en cada sistema, pero no un tiempo común a todos ellos", concepto fundamental que distingue la antigua teoría newtoniana de la nueva relativista. Considerando las distancias involucradas en un experimento *pensado*

(*experimento ideal* lo llama Collo) deduce que "las longitudes de los cuerpos dependen de la velocidad que tengan respecto del sistema en que se las mide". Estas dos conclusiones llevan directamente a las ecuaciones de transformación de Lorentz, las cuales, dadas las coordenadas espaciotemporales de un evento en un sistema de coordenadas, permiten calcularlas en cualquier otro sistema. En una larga nota al pié incluye una deducción elemental de las ecuaciones de Lorentz dada por Einstein en la obra de divulgación antes citada.

Volviendo a la idea de "escalas" en el desarrollo de la física, Collo señala que las leyes de transformación de Lorentz se reducen a las transformaciones de Galileo cuando la velocidad entre los sistemas de coordenadas es mucho menor que la velocidad de la luz: "la mecánica relativista introducirá una modificación en las relaciones de la mecánica clásica especialmente cuando se estudien las leyes de movimientos muy rápidos", aclarando que esto sucede, por ejemplo, en el caso del movimiento de los electrones[xliii]. Pasa luego a dar un sentido claro al resultado de las experiencias de Fizeau mostrando que la nueva teoría relaciona íntimamente dos principios de la física clásica: el de conservación de la masa y el de conservación de la energía.

A continuación se ocupa de las consecuencias cinemáticas y dinámicas de la teoría de la relatividad, es decir, su impacto sobre las nociones de movimiento y fuerza, formulando la famosa relación $E = mc^2$, donde $m$ es la masa relativista de la partícula en movimiento (que, en notación moderna, se relaciona con la masa en reposo a través de la relación $m = \gamma\, m_0$, donde $\gamma$ depende de la velocidad de la partícula y tiende a infinito a medida que esta velocidad se aproxima a $c$, la velocidad de la luz). El autor entonces señala que "los dos principios, de conservación de la masa y conservación de la energía [...] se refunden en uno sólo para la mecánica relativista", un cambio conceptual fundamental, pues "la masa que fue tenida siempre como absoluta (constante) es un valor relativo, dependiente de la velocidad del cuerpo".

Collo concluye su artículo con algunas consideraciones sobre la representación geométrica que Minkowski propuso en 1908 para las transformaciones de Lorentz, que permite pasar de un sistema espaciotemporal a otro, cuyas coordenadas estén definidas por la transformación de Lorentz, mediante una simple rotación angular.

**La relatividad general**

En el número siguiente del *Boletín del Centro Naval*, correspondiente a los meses de noviembre-diciembre de 1923, apareció un artículo denso de Isnardi, que ocupa treinta y siete páginas[xliv]. En este se desarrollan elementos teóricos necesarios para llegar a la teoría general de la relatividad, incluyendo los conceptos de masa gravitacional e inerte (o inercial) y el principio de equivalencia.

En la primera parte de su contribución se ocupa de las predicciones de la nueva teoría en lo referente a la deflexión de la luz en presencia de un campo gravitacional homogéneo y al corrimiento al rojo gravitacional resultante de la propagación de la luz en un campo inhomogéneo, como el del Sol. En la segunda parte calcula explícitamente las geodésicas del espaciotiempo de Schwarzschild y obtiene tanto el clásico valor de 43 segundos de arco por siglo para el avance anómalo del perihelio de la órbita del planeta Mercurio como la deflexión de 1,74 segundos de arco en los rayos de luz de las estrellas del fondo del cielo que pasan en cercanías del limbo solar.

Isnardi comienza con una discusión sobre las nociones de "reposo" y de "movimiento" absolutos. La primera había perdido toda realidad física con la introducción del principio especial de la relatividad de 1905 mientras que la segunda quedaba aún vigente ya que, incluso en la mecánica newtoniana en experiencias como la del péndulo de Foucault y en otras análogas en las que intervienen las llamadas fuerzas de inercia, se admite el movimiento absoluto del sistema de referencia. Resurgía así el *espacio absoluto*, que se creía ya olvidado por la física: no se podría ya más constatar el *reposo* del sistema (por el principio de relatividad), pero sí se podría constatar su movimiento cuando este sistema fuese acelerado (un movimiento rotatorio, por ejemplo). Ese estado de cosas era claramente insatisfactorio.

La relatividad restringida había dejado de lado a la interacción gravitatoria, pero unos pocos años más tarde Einstein propuso la equivalencia entre las fuerzas gravitatorias y las fuerzas de inercia en base a la conocida igualdad entre las masas gravitacional e inercial, es decir, el hecho de que el movimiento de los cuerpos en un campo gravitatorio es independiente de sus propiedades o, de acuerdo con Galileo, que "todos los cuerpos caen con la misma aceleración $g$". El objeto era unificar el problema y buscar una nueva teoría de la gravitación.

Isnardi discutió estos temas haciendo referencia a los libros de Arthur Erich Haas[xlv], de Eddington[xlvi], de von Laue[xlvii], de Hertz[xlviii], de Ludwig Boltzmann[xlix], de August Föppl[l] y, muy especialmente, a las monografías recopiladas en la obra de Lorentz, Einstein, Minkowski[li]. De este último libro extrae la siguiente afirmación, atribuida a Einstein: "Este hecho, de la igual caída de todos los cuerpos en un campo gravitacional, es uno de los más generales que nos ofrece la observación de la naturaleza; y, sin embargo, esta ley no ocupa ningún lugar en los fundamentos teóricos de nuestra imagen física del mundo". Isnardi explica entonces cómo Einstein fue capaz de derivar de estos conceptos el llamado *Principio de Equivalencia*, eslabón fundamental en la cadena de desarrollos teóricos que lo llevaron a formular la teoría general de la relatividad.

Enuncia ese Principio de la siguiente manera: "Para la descripción de los fenómenos mecánicos, son equivalentes un sistema inercial supuesto fijo en un campo gravitacional homogéneo, y un sistema libre de campo gravitacional, pero animado de movimiento uniformemente acelerado con respecto a un sistema inercial" y a continuación menciona el clásico ejemplo del ascensor de Einstein: "el caso de un ascensor que cayera libremente; para un observador en su interior los cuerpos carecerían de peso, y valdría la ley de inercia". En otras palabras, si estudiamos fenómenos mecánicos en un sistema inercial en presencia de un campo gravitacional homogéneo, cuando referimos nuestra descripción a un sistema de referencia que está animado de una aceleración constante igual al campo gravitacional, estos fenómenos se producen en este sistema como si se tratase de un sistema sin campo gravitacional.

No deja de lado las opiniones críticas a este Principio enunciadas por autores como Philipp Lenard[lii] y da respuestas a cada una de ellas basándose, entre otros, en el libro de Max Born[liii]. En una comparación entre la teoría de Newton y la de Einstein, de la que Isnardi se hace eco, indica que la teoría de la relatividad no intenta mostrar que la física de Newton es *ilógica*, mientras que la relatividad es *lógica*, sino que ambas son posibles, pero para que la primera tenga validez sería necesario poder "constatar movimientos absolutos; aún cuando el espacio absoluto conservara realidad metafísica". Sin embargo, Einstein había reiterado prudentemente en varias ocasiones que "los problemas metafísicos son ajenos a la teoría". Isnardi señala que la experiencia ha mostrado que no hay manera de detectar el espacio absoluto, es decir que éste "carece de realidad física"[liv].

A continuación Isnardi se pregunta si no sería posible extender el Principio de Equivalencia, que hasta ahora sólo se había discutido para los fenómenos mecánicos, a *todos* los fenómenos de la física, en particular para los electromagnéticos, afirmando que: "es ahí que reside la originalidad de las ideas de Einstein y la fecundidad del principio". En su riguroso trabajo, esta referencia a la *originalidad* y *fecundidad* del autor de la teoría de la relatividad es lo más cerca que Isnardi llega de dar una opinión subjetiva.

Siguiendo el libro de Alexander von Brill[lv], menciona que extendiendo este Principio se puede llegar, no a una teoría *explicativa* de los fenómenos producidos en un campo gravitacional, pero sí a una teoría *descriptiva* de esos fenómenos; Isnardi estima que "este es el núcleo central de la teoría general de la relatividad"[lvi].

Aprovechando el Principio de Equivalencia es posible deducir algunas consecuencias observables de cierto interés que, además, son "susceptibles de ser sometidas a la prueba experimental". Una de estas es la deflexión de la luz por la presencia de campos gravitacionales (homogéneos, como en el caso del ascensor, o inhomogéneos, como en el caso del Sol); a esta discusión Isnardi dedica un párrafo muy detallado junto a un par de gráficos. Plantea entonces el problema de la propagación de un rayo de luz en cercanías del Sol, donde el campo gravitatorio es suficientemente intenso como para afectar su trayectoria, y lo relaciona con un rayo que se propaga en un sistema acelerado en dirección perpendicular a la de propagación de la luz. En este último caso, un observador que se mueve con el sistema acelerado verá que el haz se curva en dirección contraria a su dirección de aceleración. Ahora bien, por el Principio de Equivalencia el sistema de este observador es localmente equivalente a un sistema de referencia no acelerado, pero ahora en presencia de un campo gravitacional (el del Sol). En consecuencia, en este último sistema de referencia la luz también se curvará y "la concavidad de la curva [estará] vuelta hacia el lado indicado por el sentido de la intensidad del campo".

En resumen, la luz sufre una deflexión por la gravitación de cuerpos masivos como el Sol. La verificación experimental de este efecto, cuya existencia Perrine, director del Observatorio de Córdoba, fue el primero en tratar de verificar en 1912, fue finalmente detectada durante el eclipse total de Sol de 1919 (el llamado Eclipse de Einstein). Esta referencia y su comentario en el artículo de Félix Aguilar[lvii] las hemos discutido separadamente[lviii].

Otra de las consecuencias del Principio de Equivalencia susceptible de ser sometida a experimentación es el llamado corrimiento al rojo gravitacional. Cuando la luz se propaga de un punto a otro de un campo gravitacional (no-uniforme), al alejarse de la fuente del campo (un cuerpo masivo) y por lo tanto disminuir el potencial gravitacional de la fuente, la luz sufre un corrimiento en frecuencia, virando hacia frecuencias menores, vale decir que la luz "se corre al rojo". Isnardi menciona que este efecto puede "someterse a la prueba experimental" mediante la comparación de las líneas espectrales del Sol o de las estrellas, con las de los mismos elementos producidas por una fuente en un laboratorio de la Tierra. Y enuncia que "aquellas aparecerán desplazadas hacia el rojo"[lix].

Discute luego la diferente marcha de los relojes cuando éstos se hallan en zonas distintas de un campo gravitacional, equiparando "cada átomo vibratorio con un reloj". En resumen, dos relojes exactamente iguales, "cuyas marchas coinciden si se los sitúa en el mismo lugar", dejarán de hacerlo si se los separa: aquel ubicado más cerca de la fuente del campo gravitacional (donde el potencial gravitacional es grande) retrasará con respecto al otro en un valor proporcional a la diferencia del potencial entre los dos lugares. Para esta discusión se basa en la obra de August Kopff[lx]. Este efecto, en general de valor ínfimo en la física cotidiana, fue solo medido por Robert Pound y Glen Rebka[lxi] casi cuatro décadas luego de que Isnardi escribiese su artículo.

El resto del artículo de Isnardi se centra en describir en cierto detalle el formalismo matemático de la teoría. Comienza con el espaciotiempo tetradimensional de Minkowski y discute la invariancia del intervalo entre dos eventos espaciotemporales (magnitud que él llama "segmento" pero en un espacio general de cuatro dimensiones) frente a transformaciones entre sistemas inerciales de referencia. Una oportuna cita de Minkowski, que Isnardi recoge, dice: "Desde hoy en adelante el espacio en sí mismo y el tiempo en sí mismo, se hundirán en las sombras y solamente una forma de unión entre ambos conservará existencia propia"[lxii].

En este punto de su exposición establece contacto con la última parte del artículo de Collo, cuando éste comenzaba a describir la representación geométrica de Minkowski para las transformaciones de Lorentz por medio de rotaciones sobre variables adecuadamente elegidas en un espacio de cuatro dimensiones. En una aguda nota señala que este espacio es inaccesible a nuestra intuición, pero que también lo es la electricidad, y sin embargo razonamos sobre ella.

Isnardi se detiene a señalar al lector que las propuestas de Minkowski, contrariamente a lo que se infiere de las obras de von Laue[lxiii] y Kopff[lxiv] no son una mera forma de representación en cuatro dimensiones sino que es precisamente su interpretación física lo que les atribuye un importante contenido conceptual; sin duda se refiere a la geometrización de la física. Esta observación tiene cierto interés si se la contrasta con una referencia que haremos más adelante a un trabajo de Einstein y Laub de 1908. Isnardi se acerca a los puntos de vista de Eddington[lxv], a quién cita, e ilustra esas transformaciones con un gráfico. Introduce enseguida el concepto de "línea de universo" y la locución de Eddington que, en su libro, llama a este espacio relativista, un "universo a 3 + 1 dimensiones", para destacar el diferente carácter de la dimensión temporal.

En la sección siguiente de su exposición Isnardi introduce geometrías más generales que la Minkowskiana y señala su relevancia para el desarrollo de la relatividad general[lxvi], dedicando una extensa sección a repasar "el origen y el valor de las nociones de la geometría Euclídea" como motivación al posterior desarrollo de las geometrías (curvas) de Bernhard Riemann[lxvii]. Hace referencia a la idea de Gauss, de usar una estrella para construir un triángulo geodésico inmenso que permita decidir si también en la geometría espacial los ángulos interiores suman 180°. Según Einstein los lados de ese triángulo pueden ser curvados por la influencia de, por ejemplo, la masa del Sol, lo que nos indicaría que la geometría del espacio no es necesariamente Euclídea. Concluye que la relatividad aconseja adoptar una geometría que se ajuste a las determinaciones experimentales, sin postular de antemano que ésta deba o no ser la geometría Euclídea: desde el ángulo de la física una geometría "no es ni más ni menos lógica, ni más ni menos experimentalmente cierta que la otra; puede ser, solamente, más o menos adecuada"[lxviii].

En la sección que sigue Isnardi deja de lado las conjeturas sobre la naturaleza de la geometría y pasa a discutir un caso en el que, aceptando el Principio de Equivalencia y la teoría especial de la relatividad, efectivamente se encuentra que la geometría de base no es la Euclídea.

En este punto de su exposición Isnardi se ve obligado a introducir algunas nociones sobre variedades n-dimensionales cuyos diferenciales obedecen a una forma cuadrática, es decir, a introducir elementos de la teoría de los espacios de Riemann y luego nociones de la teoría de la medida en esos espacios; sus referencias principales son ahora las *Oeuvres mathématiques* de Riemann de 1898, el trabajo geométrico de Gauss, republicado en la colección de clásicos de la ciencia dirigida por Wilhelm Ostwald, y las obras más específicas para las aplicaciones físicas debidas a Henri Galbrun[lxix] y a Hermann Weyl[lxx]. Da luego algunos ejemplos aclaratorios del rol que juegan los coeficientes variables de la forma cuadrática diferencial asociada con un elemento diferencial de arco en la definición de la geometría de una superficie y la posibilidad de aproximar la superficie en regiones infinitamente pequeñas utilizando trozos de planos tangentes, cuya forma cuadrática asociada tiene coeficientes constantes. Aplicando esas ideas a diferentes superficies introduce la noción de género. A continuación introduce el concepto de geodésica (que generaliza la propiedad de la línea recta como distancia mínima en el caso Euclídeo) en un espacio arbitrario, no necesariamente Euclídeo, como la curva de longitud límite, señalando que esta curva especial queda definida con independencia del sistema de coordenadas. Este es, precisamente, el formato en el cual sería deseable formular las leyes de la física en la nueva teoría einsteiniana de la gravitación.

Para no salir "fuera de los límites de este resumen" Isnardi deja de lado el uso del cálculo tensorial, capítulo especial del cálculo diferencial absoluto, que es la herramienta matemática que permite estudiar las relaciones métricas de las superficies de varias dimensiones. Isnardi remite al lector a fuentes especializadas que hace explícitas. Destaca entre ellas la parte teórica que Eddington agregó a la traducción francesa de su obra[lxxi], y hace también referencia a la obra de Gustave Juvet[lxxii] y a la obra ya citada de Galbrun[lxxiii].

En la sección siguiente se ocupa de los postulados fundamentales de la teoría general de la relatividad, recordando que la teoría especial exige que el espacio y el tiempo formen una multiplicidad (una variedad) riemanniana tetradimensional; su geometría será la del espacio de Minkowski. Más atrás había dado la forma cuadrática de sus diferenciales para el caso de sistemas inerciales exentos de gravitación[lxxiv]; el objetivo ahora es imponer a la variedad la forma cuadrática adecuada para definir sus relaciones métricas en un sistema de coordenadas arbitrario que describa el *campo gravitacional* en cada punto del sistema considerado.

Recuerda Isnardi que el objeto de la teoría es que sea posible hallar en cada punto del espaciotiempo sistemas de coordenadas localmente inerciales, donde valga la relatividad especial; es decir, se trata aquí de una reformulación del Principio de Equivalencia. Por último enuncia la llamada *ley de inercia*, que exige que la línea de universo de todo punto material libre sea "una geodésica en el espacio de Minkowski, curvado por la presencia de masas materiales". En el caso de la propagación de la luz, esta afirmación puede ser enunciada diciendo que "la línea de universo de la propagación luminosa es también una geodésica" aunque, más precisamente, se trata de una geodésica *nula* del espaciotiempo, donde el intervalo calculado a partir de la métrica vale cero. Resumiendo, señala la diferencia entre el espacio newtoniano y el relativista: mientras que el primero es "un espacio absoluto Euclídeo [con] fuerzas absolutas entre los cuerpos, en la teoría de la relatividad tenemos un espacio a priori totalmente informe, cuyas propiedades métricas y mecánicas quedan determinadas en absoluto por la presencia de masas materiales". Así, una partícula material, por su sola presencia, modifica el universo que la circunda de una manera absoluta[lxxv]. Como consecuencia de esta observación resulta que las comprobaciones más seguras de la teoría deberán basarse en el estudio de las trayectorias de puntos materiales libres o de rayos luminosos, ya que ellos permiten investigar el género del espacio.

El trabajo de Isnardi concluye con algunas aplicaciones importantes de la relatividad general, como son las que surgen de resolver las ecuaciones para una distribución de masa esféricamente simétrica, es decir, calcular las geodésicas en el espaciotiempo de una única partícula, de masa $M$, que puede ser una buena aproximación para el Sol o incluso para un agujero negro (y que hoy conocemos como la solución de Schwarzschild, aunque este nombre no es mencionado explícitamente por Isnardi). A partir de la métrica, Isnardi hace los cálculos *in extenso* y deduce dos de los efectos nuevos de la teoría de Einstein: el movimiento, o avance anómalo, del perihelio de Mercurio (en su desplazamiento alrededor de la *partícula* Sol, fuente del campo gravitacional) y la deflexión de un rayo de luz que pasa en las inmediaciones de un objeto esférico de masa $M$ (que nuevamente representa a nuestro Sol). En ambos casos da valores numéricos precisos, los que el mismo Einstein calculó a partir de su novedosa teoría de la gravitación.

**Einstein, Laub y la complejidad del espaciotiempo de Minkowski**

Más atrás nos hemos referido a Laub como un propulsor de la relatividad en la Argentina. El interés de Laub por esa teoría había surgido mucho antes de su emigración a ese país. Muy tempranamente, hacia 1906, percibió la originalidad de las ideas de Einstein y en 1907 publicó sus primeros trabajos en esa área, ocupándose de la óptica relativista de los cuerpos en movimiento[lxxvi]. Un año más tarde se vinculó directamente con Einstein y, aun sin tener una posición remunerada, se trasladó a Berna por algunos meses para trabajar bajo la supervisión de Einstein.

Posiblemente a causa de las dificultades que se le presentaron para obtener su *Habilitation*, hacia 1910 comenzó a buscar trabajo fuera de Alemania considerando entre otras posibilidades la de trasladarse a los Estados Unidos. Al no ser viable esta posibilidad Laub se puso en contacto con Bose, entonces en la Universidad de La Plata, dando a Einstein como una de sus referencias. Luego de algunas alternativas aceptó una cátedra en el Instituto de Física de aquella universidad, a la que se incorporó en 1911.

Una vez en La Plata Laub dictó un curso superior sobre la teoría de la relatividad que, si bien no fue el primero[lxxvii] sobre ese tema dictado en las Américas, como se ha sugerido, se cuenta entre los primeros y muy posiblemente fue el primer curso sobre esa teoría dictado en América Latina. Esto tuvo lugar cuando Collo e Isnardi estaban al final de sus carreras universitarias y se preparaban para dejar la Argentina e iniciar estudios avanzados en Alemania. Ese primer contacto con la relatividad, anterior a su viaje, puede haber activado el interés y la sensibilidad de Collo e Isnardi frente a aquella teoría.

Volvamos a la relación científica de Laub con Einstein en Europa y a un aspecto interesante de ese contacto. Si bien el concepto de espaciotiempo y la posibilidad de vincular las geometrías no-Euclídeas con la física había atraído la atención de diversos autores desde el giro del siglo XIX al XX, el concepto de espacio de Minkowski fue un primer y difícil intento de tratar el espacio y el tiempo como una unidad conceptual, aplicándolo a una situación concreta de la física.

Sin embargo, algunos especialistas europeos, incluyendo al mismo Einstein, exploraron la posibilidad de conciliar la teoría de la relatividad con posibles alternativas que simplificaran su exposición. Una de ellas fue tratar de *separar* de alguna manera el espacio y el tiempo, lo que traería simplificaciones considerables en la formulación de la teoría. Uno de esos primeros intentos tuvo realidad material en dos trabajos que Einstein publicó en 1908 precisamente en colaboración con Laub.

En su trabajo Einstein y Laub[lxxviii] estudiaron las teorías pre-relativistas del electromagnetismo desde el punto de vista relativista, dejando de lado el complejo formalismo espaciotemporal o tetradimensional que Minkowski había propuesto para la electrodinámica, y reemplazándolo por una versión más *elemental*. Esa versión, que no era estrictamente tetradimensional, daba la posibilidad de operar con algunas de sus componentes en sólo tres dimensiones, lo que permitía a los autores utilizar el *análisis vectorial* en lugar del más complejo *análisis tensorial* como herramienta matemática. El análisis vectorial era, por otra parte, un área que Laub dominaba con seguridad y de la que se había ocupado con anterioridad[lxxix].

Einstein y Laub utilizaron ese esquema para deducir las ecuaciones de los cuerpos en movimiento. Desde luego que en esta búsqueda se puede también encontrar, implícitamente, una sorprendente afirmación acerca de la visión que Einstein y Laub tendrían en 1908 acerca de cuál podría ser la geometría de la física. Sin embargo, poco más tarde Einstein comprendió la necesidad de abandonar ese enfoque. No obstante, algunos aspectos de ese trabajo conjunto no han sido totalmente olvidados.

Poco después de su arribo a la Argentina Laub publicó dos trabajos sobre la teoría de la relatividad, que aparecieron en *Anales* de la Sociedad Científica Argentina en 1912 y 1915. Los artículos de Laub no sólo continuaban los temas de su anterior colaboración con Einstein sino que nos informan de su estrecha relación con Einstein, quién hasta por lo menos 1915 parece haberle enviado manuscritos de sus trabajos.

En el primero de esos estudios[lxxx], que también publicó en *Physical Review*[lxxxi], Laub consideró modificaciones a la electrodinámica de Einstein propuestas por Max Abraham (el rechazo de la constancia de la velocidad de la luz) y de su análisis, en el que utiliza vectores, deduce consecuencias de interés directo para la verificación experimental de la teoría. De su segundo trabajo, publicado en 1915[lxxxii] se desprende que ha abandonado la idea de discutir la teoría con un cálculo vectorial tridimensional y una buena parte de este segundo trabajo está dedicada, precisamente, a introducir lo que él llama el *cálculo tetradimensional*.

Los trabajos de Laub sobre la teoría de la relatividad han quedado en un cierto remanso, aunque Cornelius Lanczos, discípulo y colaborador de Einstein en Berlín en la década de 1920, se ha ocupado de la colaboración científica que Laub tuvo con Einstein en 1908 en Alemania en su clásico estudio sobre los trabajos científicos de Einstein[lxxxiii]. Por su parte, Seelig[lxxxiv] y Pyenson[lxxxv] han historiado en detalle la relación personal entre esos dos científicos, mientras que Karin Reich ha observado con singular agudeza la estructura geométrica que encierra la propuesta de Einstein y Laub de 1908[lxxxvi].

Laub comprendió cabalmente que su colaboración con Einstein había sido un punto culminante en su carrera científica; esa colaboración fue un justo motivo de crédito para él durante el resto de su vida. Una vez en Argentina su anterior contacto directo con Einstein, y más tarde su pacifismo durante la Primera Guerra Mundial, jugaron un papel central en el establecimiento de relaciones estrechas entre Laub y un grupo destacado de intelectuales y políticos argentinos que lo apoyaron en su vida académica. Más tarde, ellos contribuyeron a facilitar su transición desde la docencia hacia cargos en la diplomacia argentina en Europa y, consecuentemente, a su alejamiento de la física por muchos años.

## CONCLUSIONES

A diferencia de un buen número de divulgadores contemporáneos de la teoría de la relatividad, Collo e Isnardi se identificaron claramente con las ideas de Einstein. En 1922 estos físicos tenían un dominio técnico considerablemente más refinado y fluido de esa teoría que la gran mayoría de sus colegas argentinos. En su trabajo no encontramos referencias anecdóticas o biográficas sobre Einstein o sobre sus colegas, las cuales eran frecuentes entonces en sectores amplios de la prensa relativista, tanto en la Argentina como fuera de ella. El trabajo de Collo e Isnardi se enfoca hacia las ideas y no hacia las personalidades o hacia la historia de la teoría que consideran. Hemos señalado más atrás que *originalidad* y *fecundidad* son los adjetivos más fuertes que Isnardi usó al referirse a Einstein.

Isnardi no ocultó las limitaciones, sin duda serias, de su trabajo. En particular, el hecho de que no presupuso en el lector el dominio del análisis tensorial, una disciplina que entonces estaba totalmente fuera de la formación matemática que se daba en los cursos universitarios en la Argentina, y hubiera reducido el círculo de sus lectores a unos pocos. No buscó tampoco la ruta de las simplificaciones, al estilo del ensayo de Einstein y Laub de 1908 que hemos mencionado y que, aunque no lo cita, difícilmente pudo haber ignorado. En cambio ofreció una gama amplia de referencias graduadas donde el lector interesado podía ganar acceso al análisis tensorial y avanzar hacia un estudio riguroso de la teoría de la relatividad generalizada.

Las referencias de Collo e Isnardi de 1922-23 muestran familiaridad con la literatura contemporánea sobre el tema considerado. Aproximadamente tres cuartos de sus referencias se refieren a obras publicadas entre 1920 y 1923. Teniendo en cuenta las fechas de esas publicaciones resulta claro que Collo e Isnardi las obtuvieron y estudiaron con posterioridad a su viaje de estudios a Europa el cual, desde luego, fue hecho antes de la publicación de la teoría generalizada de la relatividad.

Otra característica interesante de los trabajos de estos dos autores es su riqueza en contenidos conceptuales y su apertura a una discusión contemporánea de los fundamentos de la física, tema que continuó preocupando a ambos, particularmente a Isnardi, a lo largo de su carrera.

Hemos visto que consideraciones muy particulares pueden haber influido en la elección del *Boletín del Centro Naval* como vehículo para la difusión del trabajo de Aguilar, Collo e Isnardi, ya que en esos años los dos últimos comenzaban a mover el centro de sus actividades docentes fuera de la Universidad de La Plata, buscando perspectivas nuevas en una combinación de tareas entre aquella, la Escuela Naval y la Universidad de Buenos Aires.

Remarquemos, finalmente, que la reunión organizada por la Sociedad Científica Argentina, la contribución de Collo e Isnardi a ella, y su publicación en el *Boletín del Centro Naval*, marcan un punto de fractura en la discusión local de la teoría de la relatividad. Esos dos físicos, junto con el astrónomo Aguilar, contribuyeron a reposicionar el discurso estrictamente científico sobre esta teoría fuera del espacio de discusión culta general, aproximándolo a un área más técnica y específica.

Hemos señalado más atrás que el impacto intelectual de la polémica anti-positivista en la Argentina alcanzó a las ciencias hacia 1920. Con él podemos asociar tanto una nueva valoración de las ciencias teóricas como interesantes cambios que se sucedieron rápidamente en el panorama de las ciencias exactas en la Argentina a principios de esa década, que no han encontrado aún una justificación clara. Hemos dado algunos ejemplos de la nueva postura de la Universidad de Buenos Aires frente a las ciencias teóricas: la contratación del matemático puro Julio Rey Pastor y la invitación del físico teórico Albert Einstein para que disertara sobre su nueva teoría en sus claustros, son dos de los más sobresalientes. Hemos señalado también que en esos años es posible detectar a nivel nacional una desaceleración en el tradicional soporte atribuido por el estado a las ciencias experimentales[lxxxvii].

El espacio natural para el desarrollo de la física teórica era, sin duda, la universidad y muy especialmente la Universidad de La Plata donde la física había ocupado hasta entonces una posición de singular privilegio. Allí surgió, por iniciativa de Gans, la primera investigación original sobre la teoría de la relatividad: la breve nota antes citada de Loedel Palumbo. Sin embargo, en la segunda mitad de la década de 1920, cuando la superación del positivismo clásico era una realidad en las capas más significativas de la cultura argentina, la comunidad platense, que era predominantemente experimentalista y que había perdido la guía de Gans, tuvo dificultades internas para afianzar un movimiento firme hacia la teoría, en paralelo con la labor experimental. Si bien la Universidad de Buenos Aires hizo esfuerzos, difícilmente podría argumentarse que el desarrollo de la teoría de la relatividad avanzó en la universidad Argentina de la década de 1920 en proporción con el esfuerzo inicialmente empleado en promoverla[lxxxviii]. Estas circunstancias permiten apreciar más cabalmente la complejidad y las dificultades con las que estuvo asociado el proceso de transmisión e inserción de teorías científicas nuevas en la Argentina de esa década.

**Notas y referencias bibliográficas**

**Alejandro Gangui** é doutor em Astrofísica (Instituto de Astronomía y Física del Espacio, Conicet y Universidad de Buenos Aires). E-mail: gangui@iafe.uba.ar.

**Eduardo L. Ortiz** é doutor em Matemática (Imperial College, Londres). E-mail: e.ortiz@imperial.ac.uk